\documentclass[final,letter,5p,times,twocolumn]{elsarticle}
\biboptions{sort&compress}

\usepackage{lineno,hyperref}

\usepackage[T1]{fontenc}
\usepackage[latin9]{inputenc}
\usepackage{amsmath}

\usepackage{graphicx}
\usepackage{comment}

\usepackage{color}

\usepackage{graphicx}
\usepackage{multirow}
\usepackage{amsmath,bm}
\usepackage{xcolor}

\newcommand{\fzero}{$f_0(980)$}
\newcommand{\kkbar}{$K\bar K$}

\newcommand{\hcf}{Hydro-Coal-Frag}

\usepackage[normalem]{ulem}  

\renewcommand\sout{\bgroup \color{red} \ULdepth=-.5ex \ULset}

\begin{document}

\title{Probing the structure of $f_{0}$(980) from the elliptic flow in p-Pb collisions at the LHC}

\author[label1]{Yili Wang}\ead{coco.wyl@pku.edu.cn}
\author[label2]{Wenbin Zhao}\ead{wenbinzhao237@gmail.com}
\author[label4]{Che Ming Ko}\ead{ko@comp.tamu.edu}
\author[label5,label6,label7]{Fengkun Guo}\ead{fkguo@itp.ac.cn}
\author[label8,label9,label7]{Ju-Jun Xie}\ead{xiejujun@impcas.ac.cn}
\author[label1,label10]{Huichao Song}\ead{huichaosong@pku.edu.cn}

\address[label1]{School of Physics, Peking University, Beijing 100871, China}
\address[label2]{Institute of Particle Physics and Key Laboratory of Quark and Lepton Physics (MOE), Central China Normal University, Wuhan, 430079, Hubei, China}
\address[label3]{Physics Department, University of California, Berkeley, California 94720, USA}
\address[label4]{Cyclotron Institute and Department of Physics and Astronomy, Texas A\&M University, College Station, Texas 77843, USA}
\address[label5]{Institute of Theoretical Physics, Chinese Academy of Sciences, Beijing 100190, China}
\address[label6]{School of Physical Sciences, University of Chinese Academy of Sciences, Beijing 100049, China}
\address[label7]{Southern Center for Nuclear-Science Theory (SCNT), Institute of Modern Physics, Chinese Academy of Sciences, Huizhou 516000, China}
\address[label8]{State Key Laboratory of Heavy Ion Science and Technology, Institute of Modern Physics, Chinese Academy of Sciences, Lanzhou 730000, China}
\address[label9]{School of Nuclear Sciences and Technology, University of Chinese Academy of Sciences, Beijing 101408, China}
\address[label10]{Center for High Energy Physics, Peking University, Beijing 100871, China}

\date{\today}

\begin{abstract}
The $f_{0}(980)$ is a light scalar meson  whose internal structure remains under debate and investigation. Assuming that the $f_0(980)$ is a $K\bar K$ molecule that can only survive at the kinetic freeze-out of the evolving bulk matter, we implement the coalescence model to study its transverse momentum ($p_T$) spectra and elliptic flow ($v_2$) in high-multiplicity p-Pb collisions at $\sqrt{s_{NN}}=5.02$ TeV.  Using the well-tuned kaon phase-space distributions from the Hydro-Coal-Frag model, our $K\bar{K}$ coalescence calculations with reasonable values for the $f_0(980)$ radius successfully reproduce the elliptic flow measured by CMS over the range $0 < p_{T} < 12$ GeV and  also agree with the $p_T$-spectra from ALICE. These results in heavy ion collisions are consistent with the $K\bar K$ molecular picture of the $f_0(980)$. We also find that the number-of-constituent scaling of $v_2$ for the $f_0(980)$ is violated in p-Pb collisions at the LHC  because most $f_0(980)$ are produced from the  coalescence of kaons  that have different momenta.  Our study demonstrates the necessity of realistic coalescence model calculations and also explains why the CMS interpretation of the $f_0(980)$ as an ordinary $q\bar q$ meson is no longer valid by interpreting the measured $v_2$  with a simple scaling formula based on the assumption of equal momentum coalescence.  The investigation also provides a novel way to explore the internal structure of light exotic hadrons that can be abundantly produced in relativistic heavy and/or light ion collisions.
\end{abstract}

\maketitle

\bigskip
\section{Introduction} \label{sec:intro}
In the conventional quark model, mesons are described as quark-antiquark pairs and baryons as three-quark states. However, Quantum Chromodynamics (QCD) also permits the existence of exotic hadrons, including glueballs, hybrid states, and multi-quark systems, which provide valuable insights into the nonperturbative strong interactions and associated color confinement. Over the past two decades, the observation of numerous exotic hadron candidates by  Belle, BaBar, BESIII, D0, CDF, CMS, LHCb, and other experiments has sparked extensive theoretical interest, making it a particularly active area of research in hadronic physics.  One of the most important issues on the many intriguing properties of these exotic hadrons is the identification of signals that can reveal their internal structures.  In particular, there remains an active debate on whether the multiquark candidates observed in experiments are hadronic molecular states, which are weakly bound by residual strong interactions, or compact multiquark states, which are tightly bound  by the confining strong interaction; see Refs.~\cite{Jaffe:2004ph,Amsler:2004ps,Klempt:2007cp,Chen:2016qju,
Esposito:2016noz,Guo:2017jvc,Olsen:2017bmm,Liu:2019zoy,Brambilla:2019esw,
Meng:2022ozq,Chen:2022asf,Mai:2022eur,Liu:2024uxn,Chen:2024eaq, Wang:2025sic, Doring:2025sgb, Hanhart:2025bun} for reviews.

Relativistic heavy-ion collisions at the Relativistic Heavy Ion Collider (RHIC) and the Large Hadron Collider (LHC)  aim to create and study the quark-gluon plasma (QGP), a form of hot QCD matter that existed in the very early universe~\cite{STAR:2005gfr,PHENIX:2004vcz,BRAHMS:2004adc,PHOBOS:2004zne,
ALICE:2022wpn,Gyulassy:2004zy,Muller:2012zq,Shuryak:2014zxa}. As the QGP cools below the phase transition temperature, it hadronizes into a wide range of stable and unstable hadrons, including exotic ones such as the X(3872) (also denoted as $\chi_{c1}$(3872)) and the $f_0(980)$~\cite{ExHIC:2010gcb,ExHIC:2011say,ExHIC:2017smd,CMS:2021znk,ALICE:2023cxn,
CMS:2023rev}. Unlike experiments in particle physics -  where cleaner backgrounds allow for precise measurements of particle masses, widths, and decay modes - heavy-ion collisions produce these identified hadrons with sufficient abundance to enable detailed studies of their momentum distributions and flow anisotropies.  These measurements can, however, provide complementary and valuable information on the properties of produced hadrons and help probe the internal structure of exotic hadrons~\cite{ExHIC:2010gcb,ExHIC:2011say,ExHIC:2017smd}.

It has been suggested that the yields of exotic hadrons in relativistic heavy-ion collisions could be significantly affected by their internal structures~\cite{ExHIC:2010gcb}.  Subsequent calculations based on the Multi-Phase Transport (AMPT) model have also predicted that the yield of the $X$(3872) in Pb+Pb collisions could differ significantly depending on whether it is an extended molecular or a compact tetraquark state~\cite{Zhang:2020dwn}.   Recently, the CMS Collaboration has reported the first observation of the X(3872) in Pb+Pb collisions~\cite{CMS:2021znk}, albeit at very large transverse momentum. However, the yields of exotic hadron candidates can be strongly affected by medium effects, the hadronization mechanism of the QGP, and their subsequent elastic and inelastic scatterings and decays in the  expanding hadronic matter~\cite{Zhang:2020dwn,Esposito:2020ywk,
Wu:2020zbx,Zhao:2020nwy,Hu:2021gdg,Chen:2021akx,Yun:2022evm,Montana:2022inz}. These processes have not been fully explored for exotic hadrons. In contrast, the anisotropic flow, as a ratio-based observable, is less sensitive to many model-specific uncertainties~\cite{Song:2010aq}, making it a more reliable observable  for probing the internal structure of some light exotic hadrons that can be produced abundantly and  also be reconstructed in relativistic heavy-ion or light-ion collisions. 

One of such  light exotic hadrons is the $f_0(980)$, whose dominant component---whether a $q\bar{q}$ meson, compact tetraquark, or $K\bar K$ molecular state---remains under debate~\cite{Jaffe:1976ig,Jaffe:1976ih,Weinstein:1982gc,Weinstein:1983gd,Barnes:1985cy, Achasov:1987ts, Weinstein:1990gu,Morgan:1990ct,  Zou:1993az, Zou:1994ea, Janssen:1994wn,Oller:1997ti,Deandrea:2000yc, Oller:2002na, Baru:2003qq, Baru:2004xg, Maiani:2004uc, Achasov:2006cq, Hanhart:2006nr, Hanhart:2007wa, Branz:2007xp,Ahmed:2020kmp} (see the chapter ``Scalar Mesons below 1 GeV'' in the Review of Particle Physics (RPP)~\cite{ParticleDataGroup:2024cfk} for a dedicated review, and Ref.~\cite{Pelaez:2025wma} for a review of light meson resonances).  Since the $f_0(980)$ couples predominantly to the $\pi\pi$ and $K\bar{K}$ channels,  its signal interferes strongly with the background containing other broad scalar resonances (see, e.g., Ref.~\cite{Ropertz:2018stk}). Depending on its production mechanism, the $f_0(980)$ can appear as a narrow peak or may even manifest as a dip structure in the $\pi\pi$ spectrum near the $K\bar{K}$ threshold (e.g., in the $J/\psi\to\omega\pi^+\pi^-$ decay~\cite{BES:2004mws})~\cite{Dong:2020hxe}. At present, the estimated pole position of the $T$-matrix for the $f_0(980)$ in RPP is $(980 \sim 1010) - i(20 \sim 35)$ MeV, while its Breit-Wigner mass and width are $(990 \pm 20)$ MeV and $10 \sim 100$ MeV, respectively~\cite{ParticleDataGroup:2024cfk}.  Due to its relatively light mass, the $f_0(980)$ can be abundantly produced in heavy-ion or light-ion collisions and can be reconstructed via its decay channel \fzero $\to \pi^+ \pi^-$, making the measurements of its transverse momentum spectra and anisotropic flow possible~\cite{ExHIC:2017smd,ALICE:2023cxn,
CMS:2023rev}. 

Recently, the ALICE Collaboration has reconstructed the $f_0(980)$ and measured its transverse momentum spectra in p-Pb collisions at $\sqrt{s_{NN}}=5.02$ TeV~\cite{ALICE:2023cxn}. Additionally, the CMS Collaboration has measured the elliptic flow $v_2$ of the $f_0(980)$ in the same collisions. Based on a simple formula for the number-of-constituent-quark (NCQ) scaling of the $v_2$ for hadrons, CMS concluded that their results support the quark-antiquark ($q\bar q$) meson hypothesis for the $f_0(980)$ and tend to rule out any exotic structures such as  the tetraquark or \kkbar\ molecular state~\cite{CMS:2023rev}. This finding contrasts with conclusions drawn from hadron physics studies.  In Ref.~\cite{Baru:2003qq}, by generalizing Weinberg's model-independent result on compositeness~\cite{Weinberg:1965zz} to unstable particles, the authors  have shown that the $K\bar K$ component in the $f_0(980)$ is of the order of 80\% or more by analyzing the spectral function of the $f_0(980)$ using the Flatt\'e parameterization. Thus, further theoretical and experimental investigations are needed to reconcile these conflicting conclusions from the two research fields.

In this paper, we present the first systematic coalescence model calculations for the spectra and elliptic flow of the $f_0(980)$ in p-Pb collisions at the LHC.
Assuming that the $f_0(980)$ is a $K\bar K$ molecule that survives only at kinetic freeze-out, our model calculations via $K+\bar{K}\to f_{0}(980)$ coalescence nicely reproduce the differential elliptic flow $v_2(p_T)$ measured by CMS over the transverse momentum range $0 < p_{T} < 12$ GeV and are also consistent with the $p_T$-spectra measured by ALICE  by using reasonable  values for the $f_0(980)$ radius. We also find that the number-of-constituent-meson (NCM) scaling of $v_2$ is violated for the $f_0(980)$ due to a large fraction of its production arising from the coalescence of $K$ and $\bar K$ with unequal momenta. Consequently, the CMS interpretation of the $f_0(980)$ as an ordinary $q\bar q$ meson is no longer reliable, since the measured $v_2$ was analyzed using a simple scaling formula based on the equal-momentum coalescence assumption. Our investigation based on a more realistic coalescence model demonstrates, however, that the measured elliptic flow of the $f_0(980)$  is consistent with it to have the $K\bar{K}$ molecular structure.

\bigskip
\section{Methodology}  \label{sec:meth}

The coalescence model is widely employed to describe light nuclei production in relativistic heavy-ion collisions \cite{Mattiello:1996gq,Chen:2003qj,
Chen:2003ava,Zhao:2018lyf, Zhao:2020irc,Zhao:2021dka}. In this framework, the probability of forming a nuclear cluster is determined by the overlap between the cluster's Wigner phase-space density and the nucleon phase-space distributions at kinetic freeze-out. Although the $f_0(980)$ is not a stable bound state but rather a resonant (or quasi-bound) state with a strong decay channel into $\pi \pi$, its mass as given in RPP~\cite{ParticleDataGroup:2024cfk} coincides with the $K\bar{K}$ threshold. We can take its mass to be below the $K\bar{K}$ threshold  with the $K\bar{K}$ component of its wave function  similar to those of a loosely bound state~\cite{Guo:2017jvc}. 
Consequently, we assume that such a molecular configuration can only survive after the kinetic freeze-out of the evolving bulk hadronic matter. Similar to the coalescence production of the loosely bound deuteron ($d$) via $p + n \to d$ at the kinetic freeze-out~\cite{Zhao:2018lyf, Zhao:2020irc}, we calculate the coalescence production of the $f_0(980)$ via $K+\bar{K}\to f_{0}(980)$ in p-Pb collisions at the LHC.

In the coalescence model, the production probability of the $f_0(980)$ is given by the overlap of its Wigner function $\rho^{W}$ with the phase-space distributions $f_{K}(\bm{x}_1, \bm{p}_1, t_1)$ and $f_{\bar{K}}(\bm{x}_2, \bm{p}_2, t_2)$ of its constituent kaon and antikaon,
 \begin{align}\label{coal}
     \frac{dN}{d^3\bm{P}} =&\, g \int p_1^\mu d^3\sigma_{1\mu}\frac{d^3\bm{p}_1}{E_1}\, p_2^\nu d^3\sigma_{2\nu}\frac{d^3\bm{p}_2}{E_2}\notag \\
&\times  f_{K}(\bm{x}_1, \bm{p}_1,t_1) \, f_{\bar{K}}(\bm{x}_2, \bm{p}_2,t_2) \notag \\
&\times \rho^{W}(\bm{x}'_1, \bm{x}'_2;\bm{p}'_1,\bm{p}'_2;t')\, \delta^3(\bm{P} - \bm{p}_1 - \bm{p}_2),
 \end{align}
where $d^3\sigma_{i\mu}$  is the infinitesimal surface element on the switching hypersurface,  $g = (2J + 1) / \prod_{i=1}^{2}(2J_i + 1)=1$ is the statistical factor for the formation of an $f_0(980)$ with angular momentum $J=0$ from $K$ and $\bar{K}$, each having spin $J_i=0$.  In the classical distributions $f_{K/\bar{K}}$ of kaons in the coalescence formula, $\bm{x}_i$ and $\bm{p}_i$ are the coordinates and momenta of the on-shell $K$ and $\bar{K}$ in the lab frame. The coordinates $\bm{x}'_i$ and momenta $\bm{p}'_i$ in the Wigner function of the produced $f_0(980)$ are obtained by Lorentz transforming the coordinates $\bm{x}_i$ and momenta $\bm{p}_i$ to its rest frame and then propagating the earlier freeze-out $K$ {at time $t_1$} or $\bar K$ {at time $t_2$} freely with a constant velocity, determined by the ratio of its momentum to energy in the rest frame of the $f_0(980)$, to the time $t'$ when the later $K$ or $\bar K$ freezes out.

Assuming a Gaussian wavefunction for the relative motion of  its constituents $K$ and $\bar K$, the  Wigner function of the $f_0(980)$ then takes the form \cite{Ko:2012lhi}:
\begin{align}
    \rho^{W}(\boldsymbol\rho,{\bf p}_\rho)=8\exp\left[-\frac{\boldsymbol\rho^2}{\sigma_{\rho}^2}-{\bf p}_\rho^2\sigma_{\rho}^2\right],
    \label{rho}
\end{align}
where the relative coordinate and momentum $(\boldsymbol\rho,{\bf p}_\rho)$ in the rest frame of the $f_0(980)$
are defined as
\begin{align}
\boldsymbol\rho=\frac{1}{\sqrt{2}}({\bf x}_1^\prime-{\bf x}_2^\prime),\quad{\bf p}_\rho=\frac{1}{\sqrt{2}}({\bf p}_1^\prime-{\bf p}_2^\prime).
\end{align}
In Eq.(\ref{rho}), the Gaussian width parameter $\sigma_{\rho}$ is related to the root-mean-square (RMS) radius $\sqrt{\langle r^2 \rangle}$ of the $f_0(980)$ by $\sigma_{\rho} = \frac{2}{\sqrt{3}}\sqrt{\langle r^2 \rangle}$. For light nuclei, their RMS radii can be extracted from experimental measurements, which is about 1.96~fm for the deuteron~\cite{HypHI:2013sxa}.  To estimate the RMS radius of the $f_0(980)$, we follow the standard approach for near-threshold hadronic molecular states outlined in Ref.~\cite{Guo:2017jvc}. For a hadronic molecule close to its constituent threshold, the characteristic size $R$ can be approximated as $R \sim 1/\gamma$, where $\gamma =\sqrt{2\mu E_B}$ is the binding momentum of the bound state. Here, $\mu = {m_1 m_2}/{(m_1 + m_2)}$ is the reduced mass of the two-body system, and $E_B = m_1 + m_2 - M$ is the binding energy, with $m_1$ and $m_2$ being the masses of the constituents and $M$ the mass of the bound state.  With the $f_0(980)$ binding energy in the range $E_B \approx 10\text{--}20 \ \mathrm{MeV}$, such estimate yields an RMS radius of the $f_0(980)$ in the range $1.0\text{--}1.5 \ \mathrm{fm}$, which  is consistent with the expectation for a spatially extended $K\bar{K}$ molecular configuration and is used in our coalescence model calculations.

For the kaon phase-space distributions $f_{K/\bar{K}}(\bm{x}, \bm{p}, t)$ in Eq. (\ref{coal}) at the kinetic freeze-out  of p-Pb collisions, we obtain them from the \hcf\ hybrid model, which incorporates different hadron production mechanisms across the full $p_T$ range. In this framework, low-$p_T$ hadrons are produced via hydrodynamics, intermediate-$p_T$ hadrons via quark coalescence, and high-$p_T$ hadrons via string fragmentation processes~\cite{Zhao:2020wcd}. 
The hadrons generated from all three mechanisms are subsequently propagated through the Ultrarelativistic Quantum Molecular Dynamics (UrQMD) model to simulate scatterings and decays  during the hadronic evolution~\cite{Bass:1998ca,Bleicher:1999xi}. The \hcf\ model employed in this study is well-tuned and has successfully reproduced the transverse momentum spectra and differential elliptic flow of identified hadrons, including kaons, over the $p_T$ range from 0 to 6 GeV in p-Pb collisions at the LHC~\cite{Zhao:2020wcd,YuanyuanWang:2023wbz}.


\section{Results and Discussion}  \label{sec:res}

\begin{figure}[th]
\center
  \includegraphics[scale=0.55]{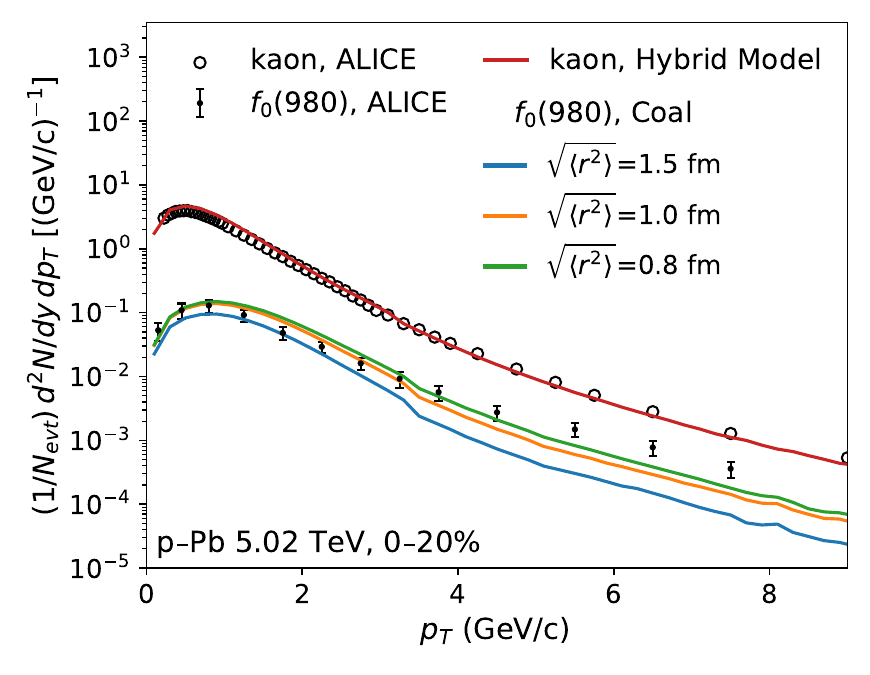}
  \caption   {(Color online) The $p_T$-spectra of \fzero\ in 0--20\% p-Pb collisions at $\sqrt{s_{NN}}$ = 5.02 TeV, calculated from the coalescence model with different RMS radius by using the phase-space distributions of kaons generated from the Hydro-Coal-Frag hybrid model.   Also plotted are the $p_T$-spectra of kaons\protect. The data are taken from the ALICE Collaboration~\cite{ALICE:2016dei, ALICE:2023cxn}.}
  \label{fig:kaon_f0_spectra}
\end{figure}

\begin{figure}[th]
\center
  \includegraphics[scale=0.55]{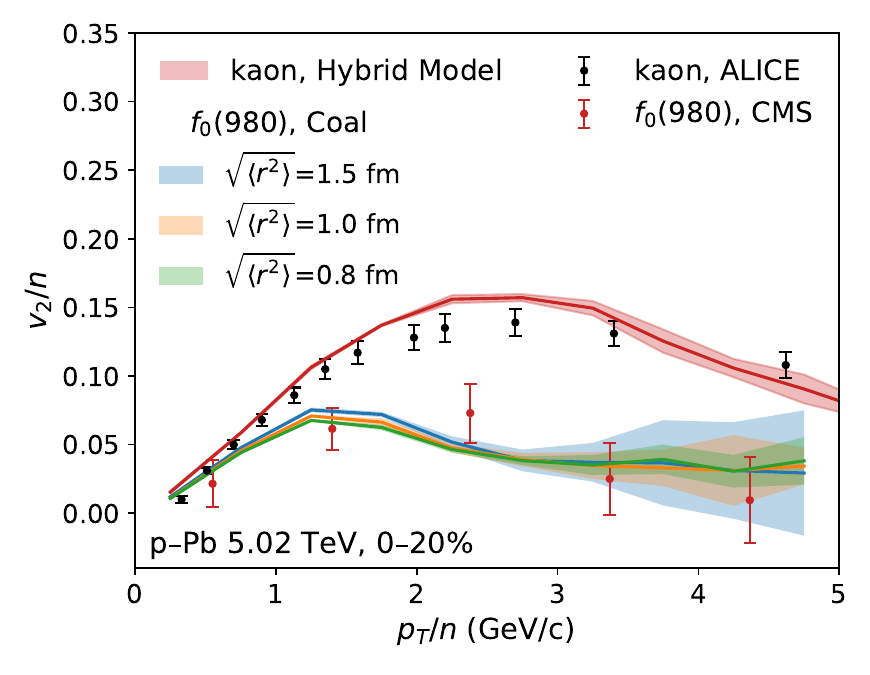}
  \caption   {(Color online) Scaled differential elliptic flow $v_2(p_T)$ of kaons (with $n = 1$) from the Hydro-Coal-Frag model and \fzero\ (with $n = 2$) from the coalescence model in 0--20\% p + Pb collisions at $\sqrt{s_{NN}}$ = 5.02 TeV. The kaon data are taken from the ALICE Collaboration~\cite{ALICE:2013snk} and the \fzero\ data  are taken from the CMS Collaboration~\cite{CMS:2023rev}.  }
  \label{fig:kaon_f0_v2pt}
\end{figure}

Figures~\ref{fig:kaon_f0_spectra} and \ref{fig:kaon_f0_v2pt}  show the transverse momentum spectra and differential elliptic flow $v_2(p_T)$ of the \fzero\ in 0--20\% p-Pb collisions at $\sqrt{s_{NN}}$ = 5.02 TeV, calculated from the coalescence model by using the phase-space distributions of kaons generated from the \hcf\ hybrid model. 
As shown in these figures, the \hcf\ model successfully fits the $p_T$ spectra and $v_2(p_T)$ of kaons over the range $0 < p_T < 6$ GeV. Using these kaon phase-space distributions, we then calculate the production of the $f_0(980)$ via $K\bar{K}\to f_{0}(980)$ coalescence. The different curves correspond to calculations with different RMS radii for the assumed $K\bar{K}$ molecular state, reflecting various spatial sizes of the $f_0(980)$.  Our model calculations reproduce the elliptic flow measured by the CMS Collaboration and are broadly consistent with the $p_T$-spectra measured by the ALICE Collaboration with reasonable  values for the $f_0(980)$ radius. The slight underestimation of the spectra at high $p_T$ of the \fzero\ in Fig.~\ref{fig:kaon_f0_spectra} is likely due to the omission of fragmentation production of the $f_0(980)$ in our calculations.  Although Fig.~\ref{fig:kaon_f0_spectra} shows that the uncertainty in the RMS radius of the $f_0(980)$ has a sizable impact on its yield, it has only a minor influence on the predicted elliptic flow $v_2$, which is a ratio-based observable. The good agreement between the $v_2$ measured by CMS and calculated by our coalescence model thus demonstrates that the flow of the $f_0(980)$ can be understood with its $K\bar{K}$ molecular structure.

\begin{figure*}[th]
\center
  \includegraphics[scale=0.55]{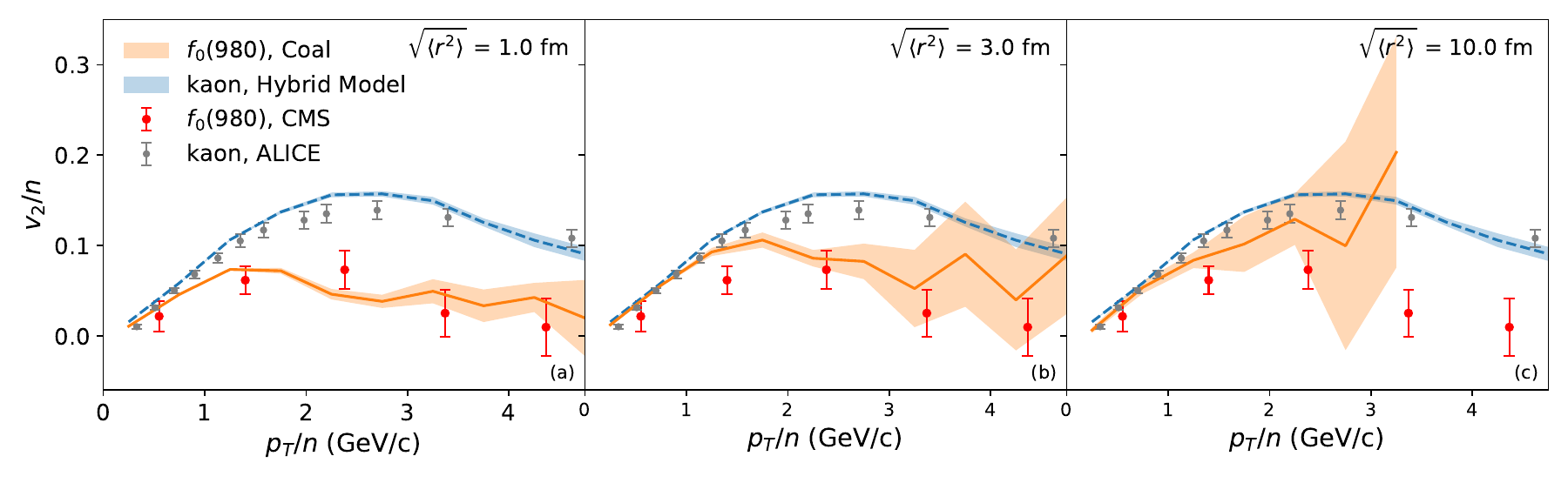}
  \caption   {(Color online) Scaling behavior for the elliptic flow between kaons ($n = 1$) and \fzero\ ($n = 2$) in 0--20\% p - Pb collisions, calculated from the Hydro-Coal-Frag \protect model for kaons and from the coalescence model with different RMS radii  \protect for the $f_0(980)$. The data are from the ALICE and CMS Collaborations \cite{ALICE:2013snk,CMS:2023rev}. }
  \label{fig:scaling}
\end{figure*}

By contrast, the CMS Collaboration has concluded that the \fzero\ was an ordinary $q\bar q$ meson by applying a simple NCQ scaling formula to the measured $v_2$~\cite{CMS:2023rev}. This simple formula reads as~\cite{Kolb:2004gi}:  
\begin{align}
v_2(p_T) / n_q \approx  v_{2,q}(p_T/n_q), 
\label{eq:ncq}
\end{align}
where $v_2$ is the elliptic flow of the produced hadron with $n_q$ constituent quarks, and $v_{2,q}$ is the elliptic flow of the constituent quark. This formula is obtained under  the assumption of equal-momentum coalescence, as further explained below. Such NCQ scaling formula can be extended to number-of-constituent (NC) scaling for an exotic hadron with a molecular structure or a light nucleus with $n$ constituent nucleons for equal-momentum coalescence.

In the coalescence production of hadrons in heavy ion and light ion collisions as given by Eq.(\ref{coal}) and used, for example, in Refs.~\cite{Zhao:2020wcd,YuanyuanWang:2023wbz}, the flow of identified hadrons produced at intermediate $p_T$ inherits the flow of its constituent quarks  due to momentum conservation~\cite{Kolb:2004gi,Fries:2003vb,Fries:2003kq,Greco:2003mm}.  With the additional assumptions that only quarks with equal momenta combine to form a hadron, one obtains the simple NCQ scaling formula given in Eq.(\ref{eq:ncq})~\footnote{Please also refer to Ref.~\cite{Guo:2016fqg} for a discussion on how NCQ scaling breaks down when the momentum scales for producing the constituent quarks are distinct.}, as used by the CMS  Collaboration. Under these assumptions, the exotic hadron $f_0(980)$ would satisfy the same scaling of $v_2$ with the number of constituent quarks $n_q=4$ for a compact tetraquark structure and with the number of constituents $n=2$~\cite{wang25} for a $K\bar K$ molecular structure.  These exotic structures are, however, ruled out when applying Eq.(\ref{eq:ncq}) to the CMS data, which prefers an ordinary $q\bar q$ meson with $n_q=2$ instead~\cite{CMS:2023rev}.

In our more realistic coalescence model calculations based on Eq.~(\ref{coal}), the Wigner function of the $f_0(980)$ in Eq.~(\ref{rho}) depends, on the other hand, on both the momentum and coordinate-space separations of its constituent $K$ and $\bar K$. Only when the parameter $\sigma_{\rho}$ becomes sufficiently large does a narrow width in momentum space, $1 / \sigma_{\rho}$, emerge. In this case, the two coalescing kaons would be widely separated in coordinate space while remaining close in momentum space, leading to a $v_2$ scaling approximately described by Eq.~{(\ref{eq:ncq})} based on the equal-momentum assumption, and the scaled flow of the $f_0(980)$ would satisfy the $n = 2$ NC hadron scaling and the $n_q = 4$ NCQ quark scaling~\cite{wang25}. However, the results shown in Fig.~\ref{fig:kaon_f0_v2pt} do not exhibit such scaling for a small but realistic value of $\sigma_{\rho}$. The large width in momentum space, $1 / \sigma_{\rho}$, enables constituents with significantly different momenta to coalesce. This invalidates the equal-momentum coalescence assumption underlying Eq.~(\ref{eq:ncq}) and breaks the NC scaling of the $f_0(980)$ $v_2$ with respect to that of  kaons or  its NCQ scaling  with respect to that of quarks. 


With increasing  RMS radius $\sqrt{\langle r^2\rangle}$ of the $f_0(980)$ in the coalescence model calculations,  the NC scaling behavior of $v_2$  is gradually restored as shown in Fig.~\ref{fig:scaling}.  However, a large $\sqrt{\langle r^2\rangle}$ around 10~fm reduces the $K\bar K$ coalescence probability  and the production of the $f_0(980)$, resulting in larger error bars in the calculated $f_0(980)$ yield. 
Also, the  large decay width of the $f_0(980)$ suggests that a radius around 10~fm is unrealistic~\footnote{Note that the $X(3872)$ with a radius of that order has a decay width much less than 1~MeV~\cite{ParticleDataGroup:2024cfk, Ji:2025hjw}. For a discussion on how molecular states of such a large size can be produced in heavy-ion collisions, see Ref.~\cite{Braaten:2024cke}.} With reasonable values of $\sqrt{\langle r^2\rangle} \approx 1\text{--}1.5$~fm, the coordinate-momentum correlations in the coalescence calculations allow two kaons with different momenta to coalesce, thereby breaking the simple NC scaling relation between kaons and the $f_0(980)$. 
This demonstrates that a  realistic coalescence calculation is necessary to describe the flow of the $f_0(980)$. Our results based on the $K\bar{K}$ molecular structure of the $f_0(980)$ agree well with the CMS flow measurements.

Note that the results shown in Figs.~\ref{fig:kaon_f0_spectra} and \ref{fig:kaon_f0_v2pt} do not include the coalescence of $\pi^+ \pi^-$ pairs in the production of the $f_0(980)$. 
Since the invariant mass of a $\pi^+ \pi^-$ pair can be much smaller than the \fzero\ mass, we assume that a $\pi^+ \pi^-$ pair can coalesce into an \fzero\ only if its invariant mass exceeds 980~MeV. Under this assumption, our model calculations show that, with  the RMS radius $\sqrt{\langle r^2\rangle}$ of the \fzero\  set to 1~fm, the  contribution from the $\pi^+ \pi^-$ channel to $f_0(980)$ production is only about $1.0 \times 10^{-6}$ of the experimentally  measured yield. This contribution further decreases to $2.0 \times 10^{-13}$ when $\sqrt{\langle r^2\rangle}$ of the \fzero\ is set to 1.5~fm. The \fzero\ elliptic flow associated with this contribution is, however,  almost indistinguishable from that obtained   from the $K+\bar{K}\to f_{0}(980)$ coalescence calculation. 

\bigskip
\section{Summary}  \label{sec:sum}

We implement  a realistic coalescence model to study the internal structure of the $f_0(980)$ meson through its elliptic flow in high-multiplicity p-Pb collisions at $\sqrt{s_{NN}} = 5.02$ TeV. In this model, the $f_0(980)$ is treated as a $K\bar{K}$ molecular state, with its internal wave function described by a Gaussian Wigner function  with its width parameter  related to the RMS radius of the produced $f_0(980)$.  The phase-space distributions of kaons at kinetic freeze-out of the evolving bulk matter are obtained from the \hcf\ model. Using these inputs and implementing $K\bar{K}$ coalescence, we calculate the production and flow of the $f_0(980)$ for a reasonable range of radius values. Our results reproduce the measured $v_2(p_T)$ from CMS over the transverse momentum range $0 < p_{T} < 12$ GeV and are broadly consistent with the measured $p_T$ spectra from ALICE. These measurements are thus consistent with the interpretation of the $f_0(980)$ as a $K\bar{K}$ molecule.

We also evaluate the number-of-constituent (NC) scaling of the elliptic flow $v_2$ for the $f_0(980)$. For a sufficiently large RMS radius around 10~fm, the coalescence dynamics approximately satisfy the equal-momentum condition, yielding a $v_2$ that obeys the simple NC scaling.  However, for a more realistic $f_0(980)$ radius around 1--1.5~fm, the large width in momentum space allows a substantial fraction of the coalescing  kaons to have different momenta, thereby breaking the naive NC scaling of $v_2$. This demonstrates the necessity of realistic coalescence model calculations for the $f_0(980)$. Consequently, the CMS interpretation of the $f_0(980)$ as an ordinary $q\bar q$ meson is no longer reliable, since the measured $v_2$ was analyzed by applying the simple scaling formula based on the equal-momentum coalescence assumption. Our calculations and investigations demonstrate that the measured elliptic flow of the $f_0(980)$ is consistent with the interpretation of the $K\bar{K}$ molecular picture for the $f_0(980)$. Also, our work  provides a novel way to explore the internal structure of other light exotic hadrons that can be abundantly produced in relativistic heavy ion and light ion collisions.


\hspace*{\fill}

\section*{Acknowledgements}
{We thank A. Gu, Y. Peng, F. Wang, H. Xu, and S.-L. Zhu for helpful discussions. This work is partly supported by the National Natural Science Foundation of China under Grant Nos.12575138, 12247107, 12125507, 12447101, 12575094, 12435007 and 12361141819 and the National Key R\&D Program of China under Grant No.2023YFA1606703. W. Z. is supported by the NSF under grant number ACI-2004571 within the JETSCAPE collaboration and by the DOE within the SURGE Collaboration. C. M. K. is supported by the DOE under Award No. DE-SC0015266. This work was performed at the Dawning Supercomputer Center (Computing Center in Xi'an), the National Supercomputer Center in Tianjin (Tianhe New Generation Supercomputer), and the High-Performance Computing Platform of Peking University.}

\appendix

\bibliographystyle{elsarticle-num}
\bibliography{new3.bib}
\end{document}